.

# THE DYNAMICS OF A CHARGED PARTICLE

Fritz Rohrlich*

*Syracuse University, Syracuse, New York 13244-1130*

Using physical arguments, I derive the *physically correct* equations of motion for a classical charged particle from the Lorentz-Abraham-Dirac equations (LAD) which are well known to be physically incorrect. Since a charged particle can classically not be a point particle because of the Coulomb field divergence, my derivation allows for that by imposing a *basic condition* on the external force. That condition ensures that the particle's finite size charge distribution looks like a point charge to the external force. Finite radius charge distributions are known not to lead to *differential* equations of motion. The present work is in agreement with the results by Spohn and by others. An example, uniform acceleration, demonstrates what the above basic condition entails. For clarity of the argument, I discuss the nonrelativistic case before the relativistic one.

.                                    PACS numbers:  03.50.-z, 03.30.+p

*Electronic address: rohrlich@syr.edu

## 1. THE PROBLEM

Since the beginning of the 20th century, the equations of motion first derived by Lorentz and Abraham [1] and rederived in covariant form by Dirac [2] have been the standard equations of motion for the relativistic dynamics of a classical charged particle. Yet, these 'LAD' equations are known to be seriously defective: they are of third order rather than second, they violate Newton's law of inertia, they have physically meaningless (runaway) solutions in the absence of an external force, and they have pre-acceleration solutions (acceleration prior to the onset of an external force) as well as post-acceleration solutions (acceleration after an external force has ceased to act.)

Despite many attempts throughout the twentieth century none have been successful in deriving a second order differential equation of motion free of all the above defects [3]. After a century of failure, Herbert Spohn succeeded in solving that problem [4]. He considered the manifold of all solutions of the LAD equation, and he identified as the *physical* sub-manifold only those solutions that have asymptotically vanishing acceleration. For this sub-manifold, all solutions can be obtained from the LAD equation by *perturbation expansion*. This is a remarkable result. It solves the century old problem of the deficiencies of the LAD equations.

The purpose of the present paper is to provide a physical argument that leads to the same equations that result from Spohn's mathematical argument. First, I provide a restriction on the external forces (equ. (1) below) that ensures that the external force cannot see the finite spread of the charge distribution. The particle will therefore appear to the external force to be a point charge. Then I show how this restriction (equation [1] below) leads from the LAD equations to the *physically correct* differential equations of motion that now replace the LAD equations. These new equations are – because of [1] – obtained by a perturbation expansion. And because of the smallness of the expansion parameter only the first order is relevant.

The divergence of the Coulomb field at its origin prevents the existence of classical *point* charges. Classical charged particles are therefore necessarily spatially extended. But extended particles lead to equations of motion that are *not* differential equations: integro-differential equations and differential-difference equations of motion for charged particles have been known for a long time. Examples were given recently by Medina [5]. If the charge distribution is restricted to the surface of a sphere, the Caldirola-Yaghjian differential-difference equations are obtained (see: Rohrlich [6]).

In order to have *differential* equations of motion, it must be possible to treat the extended charge as if it were a point when seen by the external force. This does not seem to be a problem: just use the center of mass as the representative point of the particle. This works well indeed but only for static or quasi-static forces as in gravitation. It does not work in electrodynamics where external forces can involve radiation of very short wave lengths that can "see" the finite size of the charge.

Consequently, in order to be able to use *differential* equations of motion *the external force must vary slowly enough over the size of the charge distribution* so that it will not be able to distinguish between a small but finite particle radius and a point particle. For the nonrelativistic case, this requires the force to satisfy the inequality

$$|\tau_0 \frac{d}{dt}\mathbf{F}(t)| \ll |\mathbf{F}(t)| \tag{1}$$

where $\tau_0$ is defined by (see reference [7])

$$\tau_0 = \frac{2}{3}\frac{q^2}{mc^3}. \qquad (2)$$

The time interval $\tau_0$ is very short. Its largest value occurs when $q$ and $m$ refer to an electron ($\tau_0 = 0.62 \times 10^{-23}$ sec); it is *much* shorter for other charged particles.

The force **F** is assumed to be continuous and piecewise differentiable with respect to time. The inequality (1) then states that the force as a function of $ct$ changes negligibly over a distance of the order of $c\tau_0$. That force therefore cannot see the charge as a distribution.

## 2. THE NONRELATIVISTIC EQUATIONS.

I shall discuss the consequences of (1) first for the nonrelativistic case (NR). The time rate of change of the momentum is according to the nonrelativistic limit of the LAD equation,

$$m\dot{\mathbf{v}} = \mathbf{F} + \tau_0 m\ddot{\mathbf{v}} = \mathbf{F} + \tau_0 \frac{d}{dt}(m\dot{\mathbf{v}}). \qquad (3a)$$

Multiplying equation (3) by **v** we obtain the LAD equations for the energy rate in the NR limit,

$$\frac{d}{dt}(\frac{1}{2}mv^2) = \mathbf{F}\cdot\mathbf{v} - R + \tau_0 \frac{d^2}{dt^2}(\frac{1}{2}mv^2). \qquad (3b)$$

where

$$R = m\tau_0\dot{v}^2 = \frac{2}{3}\frac{q^2}{c^3}\dot{v}^2 \qquad (4)$$

is the Larmor formula for the radiation rate.

The last terms in each of the equations (3) are Schott terms; they are characterized by time derivatives and can take on both positive and negative values. They describe internal momentum and energy rates, respectively. In this nonrelativistic approximation, the rate of emitted radiation *momentum* is too small to be included in (3a); but the rate of emitted radiation *energy* does appear (the second term on the right in (3b). The equations (3) are of *third* order thus violating Newton's first law of motion.

The correct physically meaningful equations can now be seen easily from the above. Since the time interval $\tau_0$ is so small, we would exceed the domain of classical physics and enter the domain of quantum mechanics if higher powers of $\tau_0$ than the first would be included. Taking into account the restriction (1), it follows that in this approximation, $\tau_0 m\ddot{\mathbf{v}} = \tau_0 \dot{\mathbf{F}}$, so that equation (3a) can be written as [8]:

$$m\dot{\mathbf{v}} = \mathbf{F} + \tau_0 \dot{\mathbf{F}}. \tag{5a}$$

Multiplication by **v** of this equation yields the corresponding energy rate equation

$$\frac{d}{dt}(\frac{mv^2}{2}) = \mathbf{F}\cdot\mathbf{v} + \tau_0 \dot{\mathbf{F}}\cdot\mathbf{v} = \mathbf{F}\cdot\mathbf{v} - R' + \tau_0 \frac{d}{dt}(\mathbf{F}\cdot\mathbf{v}). \tag{5b}$$

where

$$R' = \tau_0 \mathbf{F}\cdot\dot{\mathbf{v}}. \tag{6}$$

Within this approximation that ignores terms of order $\tau_0^2$ and higher, R' = R. Thus, (6) is the Larmor radiation rate.

The NR form of the LAD equations (3) thus becomes the physically correct NR equations (5). I shall call the equations (5) 'the nonrelativistic *physical* LAD equations' or 'the NR PLAD equations'. These equations are *second* order differential equations; they satisfy Newton's law of inertia, and they require that the particle acceleration vanish asymptotically since the available forces do so. But the equations must also be restricted to external forces that obey the fundamental inequality (1).

As an illustration of this new and physically correct dynamics, consider the case of a uniformly accelerated charge. This example is chosen to illustrate the importance of the condition (1). It will also confirm the emission of radiation during uniform acceleration that has been questioned repeatedly.

### 3. UNIFORM ACCELERATION

Uniform acceleration is the motion of a charged particle under a force that is (for a finite time interval) constant in the instantaneous rest frame. Relativistically, its world line in Minkowski space is a finite section of a hyperbola; hence its alternative name 'hyperbolic motion'. For NR motion, that section must also be short enough to keep the speed within the NR domain.

I assume a finite time period $2T$ during which the particle is accelerated by a constant force,

$$\mathbf{F} = m\mathbf{g} = \text{const.} \quad \text{for} \ (-T < t < T). \tag{7}$$

The requirement of continuity of $\mathbf{F}$ does not permit a step function. In fact, the fundamental requirement (1) demands a sufficiently slow rise and decline of such a force. The simplest example is motion in only one space dimension. Using $\mathbf{f} = \mathbf{F}/m$, I assume

$$f = g \quad \text{for} \ |t| < T \tag{8a}$$

$$f = 0 \quad \text{for} \ |t| > T_1 \tag{8b}$$

$$f = g \frac{T_1 - t}{T_1 - T} \quad \text{for} \ T < t < T_1 \tag{8c}$$

$$f = g \frac{T_1 + t}{T_1 - T} \quad \text{for} \ -T_1 < t < -T. \tag{8d}$$

The requirement (1) further demands that the time interval $T_1 - T$ satisfy the inequality $T_1 - T \gg \tau_0$. Ignoring that restriction leads to meaningless results. For example, such meaningless results arise in models that involve step function forces [9].

For the NR LAD equation, we find for the radiation rate (4),

$$R = \frac{2}{3}\frac{q^2}{c^3}g^2 > 0 \quad \text{for} \quad |t| < T \tag{9}$$

The NR PLAD equation (6), gives the same result since within our approximation of first order in $\tau_0$, $R' = R$. A uniformly accelerated charge *does* radiate. The last two terms in (5b) cancel: the emitted radiation rate is completely accounted for by the loss to the particle of its internal energy expressed by the Schott term. Relativistically, as can be seen in the following Section, the same cancellation takes place for the respective fourvectors of radiation energy-momentum rate and the rate of the Schott internal energy-momentum fourvector.

## 4. THE RELATIVISTIC EQUATIONS

In special relativity, equations can be written in either threevector or fourvector form. Not surprisingly, the relativistic equations are much more concise in the latter form that I shall therefore adopt. Also, to simplify the appearance of the equations and make their physical meaning more apparent, it is desirable to remove factors that clutter the equations. I shall use Gaussian units, c = 1, and the Minkowski space metric is chosen to have trace +2.

The Lorentz force, $qF^{\mu\alpha}v_\alpha$, is a special case of the general fourvector force, $F^\mu$, that acts on a particle of mass *m* and charge *q*. With the velocity $v^\mu(\tau) = =(\gamma(\tau), \gamma(\tau)v(\tau))$. The direct generalization of the NR equations (3) becomes

$$m\dot{v}^\mu = F^\mu + \tau_0 m \ddot{v}^\mu. \tag{10}$$

The differentiation in (10) is with respect to the proper time $\tau$ rather than $t$.

However, this equation is mathematically inconsistent because both $\dot{v}^\mu$ and $F^\mu$ are spacelike fourvectors, *i.e.* are perpendicular to the velocity $v^\mu$, ($\dot{v}^\mu v_\mu = 0$ and $F^\mu v_\mu = 0$) while $\ddot{v}^\mu$ is not. Therefore, it is necessary to project $\ddot{v}^\mu$ in (10) into the three-dimensional spacelike hyperplane perpendicular to the fourvelocity $v^\mu$. The projection tensor for that is $P^{\mu\nu} = \eta^{\mu\nu} + v^\mu v^\nu$ ($\eta^{\mu\nu}$ is the metric tensor). The correct generalization of equation (3) is therefore not the equation (10) but

$$m\dot{v}^\mu = F^\mu + \tau_0 m P^{\mu\alpha} \ddot{v}_\alpha. \tag{11}$$

This is exactly the relativistic LAD equation [2]; it is of *third* order. Written out more explicitly, it becomes

$$m\dot{v}^\mu = F^\mu + \tau_0 m \ddot{v}^\mu - \tau_0 m v^\mu \dot{v}^\alpha \dot{v}_\alpha. \tag{12}$$

The middle term on the right is the relativistic form of the Schott terms in (3). But now the loss due to radiation, the fourvector

$$\frac{dP^\mu}{d\tau} = R v^\mu \quad \text{where} \quad R = \tau_0 m \dot{v}^\alpha \dot{v}_\alpha \tag{13}$$

contains not only an energy loss but also a momentum loss. The radiation rate R is Lorentz invariant and is the relativistic generalization of the non-relativistic Larmor formula (4).

The transition from the LAD equation (11) to the physically meaningful PLAD equation of motion proceeds as in the NR case. Since higher orders in $\tau_0$ are negligible, $m\ddot{v}_\alpha$ in the last term of (11) can be replaced by $\dot{F}_\alpha$. This results in the PLAD equation

$$m\dot{v}^\mu = F^\mu + \tau_0 P^{\mu\alpha}\dot{F}_\alpha. \qquad (14)$$

The third order LAD equation (11) thus yields a second order PLAD equation. This is the relativistic generalization of (5).

As in the NR case, when applied to a charged particle, the equation of motion (15) must be accompanied by the fundamental restriction imposed on the external force. The relativistic generalization of (1) is (written symbolically)

$$|\tau_0 P^{\mu\nu}(\tau)\dot{F}_\nu(\tau)| << |F^\mu(\tau)|. \qquad (15)$$

I know of no case where CED gives empirically confirmed results that violate this inequality.

When considering hyperbolic motion, both LAD and PLAD give the same results relativistically as well as well as non-relativistically: the emitted radiation rates are completely accounted for by the Schott terms. More details of the relativistic case can be found in reference [10].

## 5. SUMMARY

The equations of motion (14) as restricted by (15) are the new physically correct classical relativistic differential equations of motion of a charged particle. Their NR limit is given in (5) as restricted by (1). These PLAD equations are free of the unphysical solutions present in the LAD equations. The restriction on the external force that insures that the extended charge is seen as a point charge is necessary in their derivation.

One may question the generality of this fundamental restriction. It may not hold for two reasons: either the dynamics cannot be described by a differential equation but requires a difference-differential equation (as in [6]) or some other type of equation.

Alternatively, the dynamics may be outside the classical domain and requires quantum mechanics. Failures of CED that require the use of quantum electrodynamics (QED) include scattering of radiation by a charged particle and scattering of a charged particle by a Coulomb field. Some phenomena that straddle CED and QED are discussed in the text by Jackson [11].

With the above limitations, I claim that Maxwell's equations and the new equations of motion (together with the force restriction (1) or (15)) provide a complete basis of the new classical electrodynamics. It is internally consistent, consistent with the law of inertia (LAD is not) and free of unphysical solutions.

---
I am indebted to Vladimir Hnizdo and to David Jackson for their critical reading of an earlier version and for their helpful comments.
I am indebted to Vladimir Hnizdo and to David Jackson for their critical reading of an earlier version and for their helpful comments.

---